\documentclass[prb,twocolumn,amsmath,amsfonts,floatfix,letterpaper]{revtex4}

\usepackage{graphicx}
\usepackage{ifpdf}
\usepackage[usenames]{color}

\DeclareMathOperator{\Tr}{Tr}
\DeclareMathOperator{\re}{Re}

\newcommand{\beq}[1]{\begin{equation}\label{#1}}
\newcommand{\eeq}{\end{equation}}
\newcommand{\refeq}[1]{Eq.~(\ref{#1})}

\newcommand{\beqm}[1]{\begin{multline}\label{#1}}
\newcommand{\eeqm}{\end{multline}}

\newcommand{\punc}[1]{\,{\text{#1}}}
\newcommand{\sub}[1]{_{\mathrm{#1}}}

\newcommand{\Z}{{\mathcal{Z}}}
\newcommand{\F}{{\mathcal{F}}}
\newcommand{\Ham}{{\mathcal{H}}}

\newcommand{\Lag}{{\mathcal{L}}}
\newcommand{\D}{{\mathcal{D}}}

\newcommand{\I}{\mathbf{1}} 
\newcommand{\tm}{\mathbf{t}}
\newcommand{\Gm}{\mathbf{G}}

\newcommand{\TO}{{\mathbb{T}_\tau}}

\newcommand{\ns}{^{\phantom{*}}}

\newcommand{\kv}{{\mathbf{k}}}

\newcommand{\mv}{{\vec{m}}}
\newcommand{\hv}{{\vec{h}}}
\newcommand{\hhv}{{\hat{\vec{h}}}}
\newcommand{\Sv}{\vec{S}}
\newcommand{\zerov}{{\vec{0}}}

\newcommand{\ee}{\mathrm{e}}
\newcommand{\ii}{\mathrm{i}}
\newcommand{\dd}{\mathrm{d}}

\newcommand{\kint}{\int \frac{\dd^d \kv}{(2\pi)^d}}

\newcommand{\Order}[1]{{\mathcal{O}(#1)}}

\newcommand{\ket}[1]{{|#1\rangle}}

\newcommand{\lpb}{\boldsymbol{(}}
\newcommand{\rpb}{\boldsymbol{)}}

\newcommand{\parder}[2]{\frac{\partial #1}{\partial #2}}

\ifpdf

\newcommand{\putinscaledfigure}[1]{\begin{center}\includegraphics[width=\columnwidth]{#1}\end{center}}
\else

\newcommand{\putinscaledfigure}[1]{\begin{center}\includegraphics[width=\columnwidth]{#1.eps}\end{center}}
\fi

\begin{document}

\title{Magnetic phases and transitions of the two-species Bose-Hubbard model}

\author{Stephen Powell}
\affiliation{Theoretical Physics, Oxford University, 1 Keble Road, Oxford, OX1 3NP, United Kingdom}

\begin{abstract}
A model of two-species bosons moving on the sites of a lattice is studied at nonzero temperature, focusing on magnetic order and superfluid--insulator transitions. Firstly, Landau theory is used to find the general structure of the phase diagram, and in particular to demonstrate the presence of first-order transitions and hysteresis in the vicinity of a multicritical point. Secondly, an explicit thermodynamic phase diagram is calculated using an approach based on a field-theoretical description of the Bose-Hubbard model, which incorporates the crucial effects of particle-number fluctuations. The maximum transition temperature to a magnetically ordered Mott insulator is found to be limited by the presence of the superfluid phase.
\end{abstract}

\maketitle

\section{Introduction}

Experiments with cold atoms in optical lattices have provided a new window into the physics of Mott insulator and superfluid phases, and the phase transitions between them.\cite{Jaksch,Greiner,Bloch,Catani} An important recent development in this area has been the demonstration of superexchange interactions in a spin mixture of bosonic atoms,\cite{Trotzky} suggesting that the goal of simulating magnetic Hamiltonians is within reach. The primary obstacle to this breakthrough remains the low temperatures (or entropies \cite{Bernier}) that must be reached before magnetic ordering occurs.

Perhaps the simplest possible model exhibiting the physics of magnetic ordering in itinerant systems is the Bose-Hubbard model\cite{FWGF,QPT} with two species of bosons.\cite{Kuklov,Altman,Isacsson,Numerics} In the limit of strong repulsion $U$ and weak hopping $t$, it can be described by an effective model of localized pseudospins, using a perturbative expansion in $t/U$. The coupling between these moments is induced by virtual tunneling of particles between neighboring lattice sites,\cite{Kuklov} giving a scale $J \sim t^2/U$ (see also Section~\ref{sec:PerturbationTheory}). The critical temperature for magnetic ordering is therefore strongly suppressed for $t \ll U$, and, as in the analogous case of fermions,\cite{Mathy} increasing it requires hopping strengths $t/U$ large enough that the effective local-moment description breaks down.

To describe this regime, Altman et al.\cite{Altman}\ introduced a method that builds upon mean-field theory, and includes the particle-number fluctuations that are vital to magnetic ordering. They thereby determined the zero-temperature phase diagram, including Mott insulating phases with both ferromagnetic and antiferromagnetic order, in agreement with perturbation theory,\cite{Kuklov} and also a superfluid phase, which cannot be described by an expansion in $t/U$. In recent work, S\"oyler et al.\cite{Numerics}\ used quantum Monte Carlo simulations to find the zero-temperature phase diagram, confirming most of the features found by Altman et al., and also observing phases with simultaneous superfluidity and lattice-symmetry breaking.\footnote{The approximation scheme used in this work is not capable of describing such phases and further numerical work is required to determine their stability to thermal fluctuations.}

In the present work, we address the phase diagram at nonzero temperature, with the main focus on the various types of order that are possible, including superfluid and magnetic states. The main contributions are as follows: Firstly, Landau theory is used to understand the general form of the phase diagram; most significantly, it predicts a broad region of first-order transitions and hysteresis. We then introduce a framework for the studying the phase structure of the model that uses a field-theoretical approach based on a strong-coupling expansion of the Bose-Hubbard model. We use this to present an alternative calculation of the zero-temperature phase diagram, which is mathematically equivalent to that of Altman et al.,\cite{Altman} but makes use of a quite different formalism. Finally, we demonstrate the extension to nonzero temperatures, by applying it to the calculation of the phase diagram for temperature $T>0$.

The approximation method that we use is based on the standard mean-field theory for the Bose-Hubbard model.\cite{FWGF,QPT} This can be derived by using a Hubbard-Stratonovich transformation\cite{HubbardStratonovich} to write the partition function as an integral over an auxiliary field $\psi$; the mean-field theory is given by a saddle-point approximation for this integral. In the Mott insulator, the integral is peaked at $\psi=0$, and the phase transition to the superfluid phase is signaled by a change to a nonzero expectation value of $\psi$ and hence superfluid order.

While this mean-field theory correctly predicts the phase structure of the spinless Bose-Hubbard model, it is unable to distinguish different spin-orderings within the Mott insulator. As noted by Altman et al.,\cite{Altman} this situation is similar to that encountered in frustrated magnetism,\cite{Moessner} where many different configurations have free energies that are, to a first approximation, identical. In the present case, the degeneracy of the insulating states is lifted by taking fluctuations of the on-site particle number into account.

The validity of the approach we present is controlled by the size of the fluctuations in the superfluid order parameter, which will be smaller in higher dimensions. At least in three spatial dimensions (3D), the fluctuations are expected to be relatively small, except close to the superfluid phase boundary. The approach presented here is therefore likely to be valid deep within both the insulating and superfluid phases, but necessarily breaks down in the vicinity of the phase transition to the superfluid. As usual, numerical studies are required to provide reliable results for the exact positions of the various phase boundaries.

The model that we use includes only the lowest band in the optical lattice potential, which is appropriate for the temperatures and hopping strengths that are treated. An important conclusion of this work is that the maximum transition temperature to a magnetically ordered insulator is limited by the instability to superfluidity, and to occur for hopping strengths well within the regime where a one-band model is applicable. This contrasts with the fermionic case,\cite{Mathy} where higher bands must be taken into account. We furthermore restrict throughout to the spatially homogeneous case and ignore the effects of trapping. In the presence of an external parabolic trapping potential, the results presented here apply to the Mott insulating regions in the resulting tiered `wedding cake' structure.\cite{Bloch}

In Section~\ref{sec:Model}, we introduce the model that is used throughout and review the limit where it can be described in terms of localized moments. In Section~\ref{sec:Landau}, Landau theory is applied to characterize the possible phases and transitions that the model describes. In Section~\ref{sec:FieldTheory}, we derive an approximation method based on a field-theoretical approach, and present the phase diagrams that result. We conclude in Section~\ref{sec:Conclusions} with a summary and some comments about experimental realization and detection of these phases.

\section{Model}
\label{sec:Model}

\subsection{Hamiltonian}
\label{sec:Hamiltonian}

We consider two species of bosons on a square or cubic lattice, described by the Hubbard model:
\beq{FullHamiltonian}
\Ham = -\sum_{ij,\alpha} t^\alpha_{ij} b^\dagger_{i\alpha} b_{j\alpha} + \sum_{i,\alpha\gamma} V_{\alpha\gamma} b^\dagger_{i\alpha} b^\dagger_{i\gamma} b_{i\gamma} b_{i\alpha}\punc{,}
\eeq
where sites are labeled by $i$ and $j$, and species by $\alpha,\gamma \in \{1,2\}$. The hopping matrix element $t^\alpha_{ij}$ is equal to $t_\alpha > 0$ if $i$ and $j$ are nearest-neighbors and zero otherwise. Even for $t_1 = t_2$, the Hamiltonian does not in general have $\mathrm{SU}(2)$ symmetry, instead having only $\mathrm{U}(1)\times\mathrm{U}(1)$ symmetry under independent phase rotations for the two species of bosons, corresponding to conservation of both particle numbers separately. By analogy to the $\mathrm{SU}(2)$-symmetric case, the `spin' on site $i$ can be defined in terms of the Pauli matrices $\vec{\boldsymbol{\sigma}}$ as
\beq{Spin}
\Sv _i = b^\dagger_{i\alpha} \vec{\sigma}_{\alpha\gamma} b_{i\gamma}\punc{.}
\eeq
The full spin-rotation symmetry is explicitly broken down to the set of rotations in the $x$-$y$ plane.

The calculations can be simplified considerably by restricting to the case where the intraspecies repulsion, $V_{\alpha\alpha}$, is much larger than the interspecies repulsion, $V_{\alpha\gamma}$ for $\alpha \neq \gamma$. Taking the limit of infinite intraspecies repulsion, we can describe both species by hard-core bosons and write
\beq{Hamiltonian1}
\Ham = -\sum_{ij} t^\alpha_{ij} b^\dagger_{i\alpha} b_{j\alpha} + U\sum _i n_{i1} n_{i2}\punc{.}
\eeq
This model is equivalent to that studied by S\"oyler et al.,\cite{Numerics} and has the advantage of reducing the on-site Hilbert space to that of the corresponding problem for fermions.

Our primary interest will be `magnetic' phases, where the spin degrees of freedom order, and so we restrict to the case with mean filling of one particle per site and with no population imbalance. The reduction of the problem to one of hard-core bosons introduces an additional particle-hole symmetry, and so the occupation number can be fixed with a chemical potential $\mu = U/2$.

It will sometimes be convenient to include a fictitious external `magnetic field' $\hv$, which couples to the on-site spin through a Zeeman term,
\beq{Zeeman}
\Ham _h = - \sum _i \hv_i \cdot \Sv _i\punc{,}
\eeq
and allows one to study the instability towards magnetic ordering. As we will show, this model can exhibit both ferromagnetic and (antiferromagnetic) N\'eel order, and so $\vec{h}_i$ will be taken as either uniform or staggered to describe these phases.

\subsection{Perturbation theory in $t/U$}
\label{sec:PerturbationTheory}

In the limit $t_\alpha/U \ll 1$, particle-number fluctuations are strongly suppressed and the physics is well described by an effective spin model. Restricting to the subspace where there is precisely one boson per site, the spin operator $\vec{S}_i$ describes a moment of $S = \frac{1}{2}$, and an effective Hamiltonian can be derived using perturbation theory in $t/U$, leading to\cite{Kuklov,Altman}
\beq{SpinHamiltonian}
\Ham\sub{spin} = \sum _{\langle ij\rangle} \left[ -J^\perp (S^x_i S^x_j + S^y_i S^y_j) + J^z S^z_i S^z_j\right]\punc{.}
\eeq
The coupling constants $J^\perp = 4t_1 t_2/U$ and $J^z = 2(t_1^2 + t_2^2)/U$ are both positive,\footnote{This should be contrasted with the fermionic case, where an extra minus sign from the exchange of fermions leads to antiferromagnetic $J^\perp$ as well as $J^z$.} and result from virtual tunneling processes, where a particle hops back and forth between two sites ($J^z$) or two particles with opposite spin on adjacent sites exchange places ($J^\perp$).

The phase structure in this limit is therefore described by a quantum spin-$\frac{1}{2}$ XXZ model: At zero temperature, there is a first-order transition between a ferromagnet with $\langle \vec{S}_i \rangle$ in the $x$-$y$ plane for $J^\perp \gg J^z$ and a N\'eel state with $\langle \vec{S}_i \rangle$ along the $z$ axis for $J^z \gg J^\perp$. Both give way to a paramagnetic phase at a critical temperature $T\sub{C}$ proportional to the coupling.

This perturbative calculation is only applicable in the limit of weak tunneling, and so cannot describe transitions to a superfluid state. It nonetheless provides the important insight that the tendency towards magnetic ordering is caused by the enhancement of particle-number fluctuations between magnetically aligned sites. In Section~\ref{sec:FieldTheory}, we show how this physics can be incorporated in a calculation of the phase diagram that is valid beyond the perturbative limit.

\section{Landau theory}
\label{sec:Landau}

To understand the general form of the phase diagram, and the nature of the phase transitions that are possible between the different types of order, it is useful to consider the predictions of Landau theory. We first identify order parameters to distinguish the various phases of interest, then construct an expansion for the free energy in terms of these, including all terms allowed by symmetry.

As noted above, the model is capable of both ferromagnetic order, where the spins are aligned in the $x$-$y$ plane (and which will be referred to as the `XY-ferromagnet'), and N\'eel order, with the spins aligned with the $z$ axis. We therefore define the uniform magnetization $\mv \sim \langle \Sv _i \rangle$ and the staggered magnetization $n \sim \eta _i \langle S^z_i \rangle$ (where $\eta _i = \pm 1$ on the two sublattices). Note that fixing the total particle number of the two species to be equal (and assuming the absence of phase separation\footnote{Infinite intraspecies repulsion means that real-space segregation of the two species is likely to be extremely unfavorable energetically.}) implies that $\mv$ lies in the $x$-$y$ plane, and we further assume that the N\'eel order is aligned with the $z$ axis.

As in the spinless case, the Hubbard model can also demonstrate superfluid phases, in which the phase rotation symmetry of one or both boson species is spontaneously broken, and which are described by the two superfluid order parameters $\chi _\alpha \sim \langle b_{i\alpha} \rangle$. The phase in which both are nonzero, which will be referred to simply as the superfluid, necessarily also has nonzero $\mv$, with its direction in space determined by the relative phase of $\chi _1$ and $\chi _2$.

In their numerical study of the zero-temperature phase diagram of this model, S\"oyler et al.\cite{Numerics}\ also find phases with superfluidity of only one of the two species coexisting with checkerboard density order. Such a `checkerboard superfluid' has $\chi _1 \neq 0$ (say) and $n \neq 0$, while $\chi _2$ and $\mv$ vanish.

We first treat the superfluid and ferromagnetic phases, for which the Landau theory involves the order parameters $\mv$, $\chi _1$ and $\chi _2$, and takes the general form
\begin{multline}
\label{LandauTheory1}
\Lag _{m,\chi} = r_1 \chi^*_1 \chi\ns _1 + r_2 \chi^*_2 \chi\ns_2 + u_{11} (\chi^*_1 \chi\ns_1)^2 + u_{22} (\chi^*_2 \chi\ns_2)^2\\
+ u_{12} (\chi^*_1 \chi\ns_1)(\chi^*_2 \chi\ns_2) + R |\mv|^2 + U(|\mv|^2)^2 \\+ D (m^z)^2 + \gamma \mv \cdot (\chi^\dagger \vec{\sigma} \chi) + \cdots\punc{.}
\end{multline}
The term with coefficient $D > 0$ constrains $\mv$ to lie in the $x$-$y$ plane, the last term is the lowest-order coupling allowed between the magnetic and superfluid orders, and the ellipsis represents further terms of higher order. According to Landau theory, the phase structure can be determined by minimizing $\Lag_{m,\chi}$ as a function of the parameters appearing in this expression, which are in turn (unknown) functions of the physical parameters appearing in the Hamiltonian.

In the superfluid phase, both $\chi _1$ and $\chi _2$ become nonzero, with the relative magnitude determined by the ratios between the coefficients $r_{1,2}$ and $u_{11,22,12}$. Their relative phase remains arbitrary, but is fixed once the direction of $\mv$ in spin-space is determined: for $\gamma > 0$, $\mv$ and $\chi^\dagger \vec{\sigma} \chi$ are parallel, while for $\gamma < 0$, they are antiparallel. The minimum of $\Lag_{m,\chi}$ can therefore be found from the simplified expression
\beq{LandauTheory2}
\Lag _{m,\chi}' = a \chi^2 + b \chi^4 + A m^2 + B m^4 + \alpha m \chi^2 + \cdots\punc{,}
\eeq
where the vector structure and complex phase of $\mv$ and $\chi _\alpha$ have been eliminated by requiring that $\Lag_{m,\chi}$ be minimized: the minimum always occurs when the vectors $\mv$ and $\chi^\dagger \vec{\sigma} \chi$ are aligned (and lie in the $x$-$y$ plane). Note that $\alpha < 0$, without loss of generality, since this minimization requirement fixes the relative orientation of $\mv$ and $\chi^\dagger \vec{\sigma} \chi$.

\begin{figure}
\putinscaledfigure{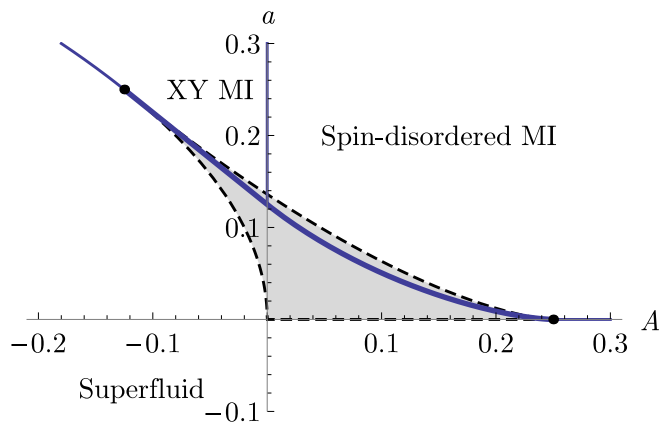}
\caption{\label{landauPlot}(Color online) Phase diagram predicted by Landau theory, determined by minimizing the free-energy function $\Lag _{m,\chi}'$ given in \refeq{LandauTheory2}, as a function of $A$ and $a$. (The parameters $B = b = -\alpha = 1$ are fixed without loss of generality, and higher-order terms are set to zero.) The phases included are the spin-disordered Mott insulator (MI), the XY-ferromagnetically ordered MI, labeled `XY MI', and the superfluid, which also has XY-ferromagnetic order. The solid lines dividing the two MI phases and dividing the superfluid from the MI phases for $A > \frac{1}{4}$ and $A < -\frac{1}{8}$ are continuous transitions. The thick line shows a first-order transition between the superfluid and MI phases, which is surrounded by a hysteretic region, shown shaded and surrounded by dashed lines.}
\end{figure}

The phase structure implied by this Landau action is shown in Figure~\ref{landauPlot}, as a function of the parameters $a$ and $A$, with the assumption that $B$ and $b$ are positive. The phase transitions separating the three phases are indicated with solid lines, which are thin for continuous transitions and thick for first-order. The shading indicates the region of hysteresis, where two local minima of $\Lag_{m,\chi}'$ exist and the stable phase is determined by the global minimum.

In the plot, the coefficients $b = B = -\alpha = 1$ have been fixed without loss of generality (assuming positive $b$ and $B$; by rescaling $m$, $\chi$ and the overall scale of $\Lag _{m,\chi}'$), and the coefficients of the remaining terms, represented by the ellipsis in \refeq{LandauTheory2}, have been set to zero. As usual in Landau theory, the positions of the continuous transitions depend only on the coefficients of the first few terms in the expansion, but the precise shape of the hysteretic region and the position of the first-order line also depend on the higher-order coefficients. (For $B$ or $b$ negative, further terms in the expansion become important and first-order boundaries will extend over a larger region of the diagram.)

The following general conclusions of Landau theory are illustrated in Figure~\ref{landauPlot}: Firstly, the three phases, spin-disordered Mott insulator (MI), XY-ferromagnetic MI, and superfluid, meet at a point. Secondly, the transition between the two insulating phases, across which $\mv$ becomes nonzero, is continuous (assuming that the quartic coefficient $B$ is positive). Finally, the transitions from the insulating phases to the superfluid, across which $\chi _1$ and $\chi _2$ become nonzero, are of first order in a region surrounding the point where all three phases meet, but can be continuous elsewhere. In Section~\ref{sec:FieldTheory}, we will develop an approximate treatment of the microscopic model and show that its conclusions are in agreement with those presented here (see, in particular, Figure~\ref{Phases3dtT}).

It is straightforward to include the N\'eel phase in the above analysis, by considering couplings of the order parameter $n$ to $\mv$ and $\chi _\alpha$: Due to the presence of the staggering factor $\eta _i$, the only allowed terms involve $n^2$, a scalar under the symmetry group. As a consequence, Landau theory predicts that a direct transition from a N\'eel phase, with $n\neq 0$ and $\mv = \zerov$, to either the superfluid or XY-ferromagnet must be of first order. This prediction is again in agreement with the results of Section~\ref{sec:FieldTheory} (Figures~\ref{Phases3d8} and \ref{Phases3dfixedT}).

Finally, the checkerboard superfluid, with $\chi _1 \neq 0$ and $n \neq 0$, can be connected by a continuous transition to the N\'eel phase, equivalent to the standard superfluid--insulator transition of a single species of boson. A direct transition to the uniform superfluid, where both $\chi _1$ and $\chi _2$ are nonzero, is necessarily of first order. These observations are consistent with the results of quantum Monte Carlo simulations.\cite{Numerics}

\section{Field-theoretical approach}
\label{sec:FieldTheory}

As seen in Section \ref{sec:PerturbationTheory}, perturbation theory in $t_\alpha/U$ predicts a critical temperature $T\sub{C}$ for magnetic ordering proportional to $t_\alpha^2/U$. Maximizing $T\sub{C}$ therefore requires increasing the hopping beyond the limit where this leading order result is valid. While it is possible to continue the expansion in $t_\alpha/U$ to higher order, generating further couplings between spins, such an approach cannot describe the superfluid phase. We instead use an approximation based on the mean-field theory that has been to applied the spinless case.\cite{FWGF,QPT}

\subsection{Partition function}

The thermodynamic properties are described in terms of the partition function, defined by $\Z = \Tr \ee^{-\beta \Ham}$, where the trace is over the full Hilbert space for all sites, and $\beta = 1/T$ is the inverse temperature. (Here and throughout, we set $k\sub{B} = 1$.) The mean-field theory for the Bose-Hubbard model can be derived starting from a Hubbard-Stratonovich transformation,\cite{HubbardStratonovich,FWGF,QPT} which rewrites $\Z$ in terms of an integral over a complex field $\psi _\alpha$:
\beq{PartitionFunction2}
\Z = \frac{\displaystyle\int \D^2\psi \: \exp -\!\left[\int _0^\beta \dd \tau\, \psi^\dagger(\tau) \tm^{-1} \psi(\tau) + \Omega[\psi]\right]}{\displaystyle\int \D^2\psi \: \exp -\left[\int _0^\beta \dd \tau\, \psi^\dagger(\tau) \tm^{-1} \psi(\tau)\right]}\punc{.}
\eeq
The auxiliary field $\psi _\alpha$ has the same vector structure as the boson operator $b_{i\alpha}$, and we use the shorthand
\beq{SumShorthand}
\psi^\dagger(\tau) \tm^{-1} \psi(\tau) = \sum_{ij,\alpha} \psi _{i\alpha}^\dagger(\tau) (\tm_\alpha^{-1})_{ij} \psi _{j\alpha}(\tau)\punc{.}
\eeq

Once the hopping term has been decoupled using the Hubbard-Stratonovich transformation, the remaining terms in the Hamiltonian each act only at a single site. The effective action $\Omega$ can therefore be found in terms of the solution of a one-site problem:
\beq{Omega}
\Omega[\psi] = -\sum _i \log \Tr \TO \ee^{-\int _0^\beta \dd \tau \Ham _i\lpb\psi _{i\alpha}(\tau)\rpb}\punc{,}
\eeq
where $\TO$ denotes ordering in imaginary time and
\begin{multline}
\label{Hami}
\Ham _i(\psi _\alpha) = U n_{i1} n_{i2} - \mu(n_{i1} + n_{i2}) - \vec{h}_i \cdot b^\dagger_{i\alpha} \vec{\sigma}_{\alpha\gamma} b_{i\gamma}\\{}- \sum_{\alpha}(\psi_{\alpha}^* b_{i\alpha} + \psi _{\alpha}b^\dagger _{i\alpha})
\end{multline}
is the local effective (time-dependent) Hamiltonian.

\subsubsection{Gaussian approximation}

To reduce the functional integral in the numerator of \refeq{PartitionFunction2} to a tractable form, we approximate the effective action $\Omega[\psi]$ by an expansion up to quadratic order around its minimum. The validity of this approximation is controlled by the size of the fluctuations, which are expected to be larger closer to the superfluid transition and in fewer spatial dimensions. In particular, in the superfluid phase in 2D at nonzero temperature, the superfluid order parameter is completely eliminated by fluctuations, and this approach is not expected to be applicable. In 3D, we expect the approximation to give reasonable quantitative results, except in the region close to the transition to the superfluid, where the gap to single-particle excitations vanishes and a quadratic approximation ceases to be valid. This approach also omits fluctuations of the magnetization, which become large in the region close to the magnetic ordering transition.

For the purposes of studying the magnetic ordering in the insulator and finding the phase boundary to the superfluid, it is sufficient to expand $\Omega$ around the point where $\psi _\alpha = 0$. Using time-dependent perturbation theory, \refeq{Omega} can be expanded in powers of $\psi$ as
\begin{widetext}
\beq{OmegaExpansion}
\Omega = -\sum _i \log \Tr \ee^{-\beta \Ham _i(0)} + \frac{1}{\beta} \sum _\omega \sum _{i,\alpha\gamma} \psi _{i\alpha\omega}^* G_{i\alpha\gamma}(\ii\omega) \psi _{i\gamma\omega} + \Order{\psi}^4\punc{,}
\eeq
where\cite{vanOosten,Sengupta}
\beq{G1}
G_{i\alpha\gamma}(\ii\omega) = -\frac{\displaystyle\sum _{n} \ee^{-\beta\epsilon _{n}} \sum _{n'} \left(\frac{\langle n | b _\alpha ^\dagger | n' \rangle\langle n' | b_{\gamma} | n \rangle}{\ii \omega + \epsilon _{n'} - \epsilon _{n}}+\frac{\langle n | b _{\gamma} | n' \rangle\langle n' | b^\dagger _\alpha | n \rangle}{-\ii \omega + \epsilon _{n'} - \epsilon _{n}}\right)}{\sum _{n} \ee^{-\beta\epsilon _{n}}}
\eeq
\end{widetext}
and $\sum _\omega$ denotes a sum over all (bosonic) Matusbara frequencies $\omega$. In \refeq{G1}, the indices $n$ and $n'$ label eigenstates of the single-site Hamiltonian $\Ham _i(0)$, and $\epsilon _n$ and $\epsilon _{n'}$ are the corresponding eigenvalues. In the zero-temperature limit, the sum over $n'$ becomes a sum over excitations, in this case double and zero occupation, above the on-site ground state. (The restriction of the on-site Hilbert space to that of hard-core bosons significantly simplifies the calculation of this quantity.)

Once the expansion in \refeq{OmegaExpansion} has been truncated to quadratic order, both integrals over $\psi$ in \refeq{PartitionFunction2} are Gaussian and can be calculated to give the free energy, $\F = -\frac{1}{\beta}\log \Z$. We find $\F = \F _0 + \F _1$, with
\begin{align}
\F _0(\hv) &= -\frac{1}{\beta}\sum _i \log \Tr \ee^{-\beta \Ham _i(0)}\punc{,}\label{FreeEnergy0}\\
\F _1(\hv) &= \frac{1}{\beta} \sum _\omega \log \det \left[ \I + \tm \Gm(\ii \omega) \right]\punc{,}
\label{FreeEnergy1}
\end{align}
where $\tm$ and $\Gm$ are matrices in sites and flavor indices. Note that the dependence on the external field $\hv$ is due to the Zeeman term in $\Ham _i$ and the resulting dependence on $\hv$ of the eigenstates and eigenvalues appearing in \refeq{G1}.

The saddle-point contribution to the free energy, $\F_0$, has no dependence on the hopping $t_\alpha$ and only contains information about the on-site state. The Gaussian fluctuations about this saddle point give $\F_1$, which takes into account the particle-number fluctuations that are crucial for describing the magnetic ordering transitions. In the case where $\vec{h}_i$ is uniform, $\F_1$ can be written as a sum over eigenvectors of $\tm$, labeled by momentum $\kv$; in the thermodynamic limit this becomes a $d$-dimensional integral over $\kv$. When the applied field is staggered (such as when considering N\'eel ordering), there is mixing between momenta separated by the appropriate reciprocal lattice vector, but the same approach can be used.

The Matsubara sum in \refeq{FreeEnergy1} converges because the determinant tends (sufficiently rapidly) to $1$ for large frequencies. Since $\Gm(z)$ and hence $\det \left[ \I + \tm \Gm(z) \right]$ are rational functions, the sum can be calculated using the result that for any polynomials $P$ and $Q$ such that $\lim _{|z|\rightarrow \infty} \frac{P(z)}{Q(z)} = 1$,
\begin{multline}
\label{MatsubaraSum}
\frac{1}{\beta} \sum _\omega \re \log \frac{P(\ii\omega)}{Q(\ii\omega)} = \frac{1}{2\beta} \left[ \sum _{z_P} \log(\cosh \beta z_P - 1)\right.\\
\left.- \sum _{z_Q} \log(\cosh \beta z_Q - 1) \right]\punc{,}
\end{multline}
where $z_P$ and $z_Q$ are the zeros of $P(z)$ and $Q(z)$.

The calculation of the free energy $\F$, as a function of the applied field, the temperature and the parameters in the Hamiltonian, is therefore reduced to the problem of calculating the matrix $\Gm$ and integrating over $\kv$.

\subsubsection{Single-particle gap}
\label{sec:Gap}

\begin{figure}
\putinscaledfigure{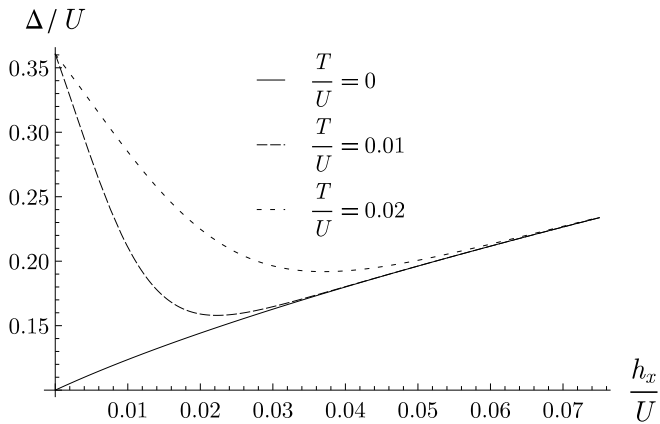}
\caption{\label{GapPlot}Single-particle gap $\Delta$, in units of the on-site interaction strength $U$, as a function of (uniform) applied magnetic field $h_x$, for three different temperatures. The hopping strengths are $t_1 = t_2 = 0.04 U$ and the lattice coordination is $z=6$. For $h_x < T$, the gap decreases rapidly with $h_x$ as the spins become ordered, enhancing particle-number fluctuations between neighboring sites. This fluctuation effect can lead to spontaneous magnetic ordering in the absence of a field and is captured by including Gaussian corrections in the calculations of Section~\ref{sec:FieldTheory}. For $h_x \gg T$, the magnetization is saturated and the Zeeman term causes the gap to increase linearly with $|h_x|$.}
\end{figure}
The perturbative calculation in Section~\ref{sec:PerturbationTheory} shows that magnetic ordering results from the enhancement of hopping between sites with aligned spins. One therefore expects that the gap $\Delta$ to single-particle excitations should depend on the magnetization, and that capturing this effect is crucial to describing the magnetic phases. By inverting the Hubbard-Stratonovich transformation, the propagator of the bosons $b_{i\alpha}$ can be related\cite{Sengupta} to that of the field $\psi _\alpha$, and is given within our approximation scheme by $[\Gm(\ii \omega)^{-1} + \tm]^{-1}$. The single-particle excitations are described by the poles of this expression for real $z = \ii \omega$, and the gap $\Delta$ is given by the smallest excitation energy. Figure~\ref{GapPlot} shows a plot of this quantity in the insulating phase, in the presence of a uniform applied field in the $x$ direction.

For the temperatures shown in the plot, thermal particle-number fluctuations are strongly suppressed, and the lowest-energy on-site configuration has a single particle. For $h_x \ll T$, this particle can have any spin orientation, while for $h_x \gg T$, the spin is fixed to point along the $x$ axis. For $h_x$ considerably smaller than $T$, increasing the field makes the on-site configuration more spin-polarized, allowing for more coherent particle-number fluctuations, and decreasing the gap. This continues until the applied field is somewhat larger than the temperature and the on-site state is essentially completely spin polarized. The gap then increases with increasing field, since the on-site energy cost of particle-number fluctuations grows linearly with $|\vec{h}|$.

\subsection{Magnetic ordering}

The phase boundaries for magnetic ordering can be found using the expression for the free energy $\F(\hv)$ given in Eqs.~\ref{FreeEnergy0} and \ref{FreeEnergy1}. Different approaches are required for zero and nonzero temperatures.

\subsubsection{Zero temperature}

In the limit $\beta\rightarrow\infty$, the only term that contributes to the sum over $n$ in \refeq{G1} is where $\ket{n}$ is the ground state of the on-site Hamiltonian $\Ham _i(0)$. There is therefore a discontinuity at zero applied field, because $\Ham _i(0)$ then has a two-fold degenerate ground state. This degeneracy results from the spin degree of freedom, and is split by an infinitesimal $\vec{h}$. The limit of the free energy $\F$ as $\vec{h}\rightarrow 0$ therefore depends on the direction in spin-space, and on the configuration of $\vec{h}_i$ in real space. The ground state is found by minimizing $\F$ with respect to the direction of $\vec{h}$ in the limit that its magnitude goes to zero.

For vanishing applied field, $\F_1$ becomes
\begin{multline}
\label{FzeroTuniform}
\F_1^{\mathrm{unif.}} = \kint \left(-\frac{U}{2} \right.\\
\left.{}+ \frac{1}{2}\sqrt{(t^1_\kv+t^2_\kv)^2\cos^2 \theta - 2(t^1_\kv+t^2_\kv)U + U^2}\right)\punc{,}
\end{multline}
for a uniform field applied at an angle $\theta$ from the $z$ axis. The free energy is minimized by $\theta = \frac{\pi}{2}$, i.e., by taking the applied field to lie in the $x$-$y$ plane, in agreement with the results of perturbation theory in Section \ref{sec:PerturbationTheory}. For $\theta = \frac{\pi}{2}$, the free energy reduces to
\beq{FzeroTuniform2}
\F_1^{\mathrm{unif.}} = \kint \left( -\frac{U}{2} + \frac{1}{2}\sqrt{U^2 - 2(t^1_\kv+t^2_\kv)U} \right)\punc{.}
\eeq
The corresponding expression for a staggered field is minimized by a field along the $z$ axis, $\theta = 0$, for which the free energy is given by
\begin{multline}
\label{FzeroTstaggered}
\F_1^{\mathrm{stag.}} = \kint \left(-\frac{U}{2} + \frac{1}{4}\sqrt{U^2 - 4t^1_\kv}\right. \\
\left.{}+ \frac{1}{4}\sqrt{U^2 - 4t^2_\kv}\right)\punc{.}
\end{multline}

These expressions for $\F$ agree with those given by Altman et al.,\cite{Altman} in particular their Eqs.~(33) and (41). Expanding in powers of the hopping,\cite{Altman} one reproduces the results of perturbation theory, given above in Section \ref{sec:PerturbationTheory}.

\subsubsection{Nonzero temperature}

For $\beta < \infty$, $\Gm$ is a continuous function of the applied field, and the previous approach no longer applies. To determine where magnetic ordering takes place, we find the effective potential $\Phi$ as a function of the magnetization $\vec{m}$. First, we define $\hat{\vec{h}}(\vec{m})$ as the solution of
\beq{hhDefine}
{\left.-\parder{\F}{\hv}\right|}_{\hhv(\mv)} = \mv\punc{.}
\eeq
The effective potential $\Phi(\mv)$ is then given by the Legendre transform of $\F(\hv)$:
\beq{Legendre}
\Phi(\mv) = \mv\cdot\hhv(\mv) + \F\lpb\hhv(\mv)\rpb\punc{.}
\eeq

The expression for the free energy, $\F = \F_0 + \F_1$, can be viewed as the first two terms in an expansion in the order of the fluctuations, and we similarly seek $\Phi(\mv)$ as a series in fluctuations. For this we require only the zero-order expression $\hhv_0$, satisfying $-\F_0'\lpb\hhv_0(\mv)\rpb = \mv$: the expression for $\Phi(\mv)$ is given by
\beq{Phi}
\Phi _0(\mv) + \Phi _1(\mv) = \mv\cdot\hhv_0(\mv) + \F_0\lpb\hhv_0(\mv)\rpb + \F_1\lpb\hhv_0(\mv)\rpb\punc{.}
\eeq

This result for $\Phi$ differs from one that would be obtained by performing the Legendre transformation directly with $\F = \F_0 + \F_1$, and is equivalent to an RPA-like summation. For example, (the exact result for) the quadratic coefficient in a series expansion of $\Phi$ is given by $\Phi''(\zerov) = -[\F''(\zerov)]^{-1} = \chi^{-1}$, the reciprocal of the magnetic susceptibility. The expression in \refeq{Phi}, which gives $\Phi''(\mv)$ consistently in powers of the fluctuations, is therefore equivalent to resumming a geometric progression:
\beq{Dyson}
\Phi''(\zerov) = (\chi _0 + \chi _1)^{-1} \simeq \left(\frac{\chi_0}{1 - \frac{\chi_1}{\chi_0}}\right)^{-1} = \chi_0^{-1} - \frac{\chi _1}{\chi _0^2}\punc{.}
\eeq
As in the familiar case of the Stoner criterion, the phase transition to a magnetic state occurs when the susceptibility diverges, and the `correction' has the same magnitude as the leading order result.

To find the magnetic phase structure within the Mott insulator, we therefore calculate $\Phi$ according to \refeq{Phi} and minimize it with respect to the magnetization $\mv$. If the minimum occurs for zero magnetization, then the system is magnetically disordered. If it occurs for uniform nonzero $\mv$, the system is ferromagnetically ordered, while if it occurs for staggered $\mv$, the system is antiferromagnetically ordered.

\subsection{Superfluid--insulator transitions}

As described in Section \ref{sec:Gap}, the energies of single-particle excitations can be determined by locating the poles of the bosonic propagator. At the transition from the insulator to the superfluid, the gap vanishes and the Gaussian approximation scheme breaks down. As in the spinless case,\cite{FWGF,QPT,vanOosten} it is nonetheless possible to estimate the position of the phase boundary as the point where the lowest-order estimate for the gap vanishes.

In the present case, however, it should be noted that the single-particle gap is dependent on the magnetization, as illustrated in Figure~\ref{GapPlot}, and it is therefore possible for the gap to vanish first at a value of magnetization away from the minimum of the effective action. In the region where the two hopping strengths $t_1$ and $t_2$ are of similar magnitude, and ferromagnetic ordering is favored, the gap is in fact found to decrease with increasing magnetization, as expected from the analysis of Section \ref{sec:Landau}. As predicted by Landau theory, this can lead to a first-order transition, where the appearance of a superfluid order parameter is accompanied by a discontinuous increase in the magnetization.

\subsection{Phase diagrams}

\begin{figure}
\putinscaledfigure{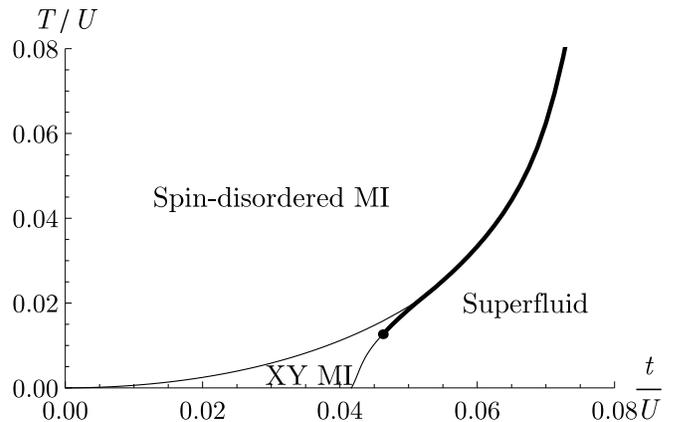}
\caption{\label{Phases3dtT}Phase diagram on the simple cubic lattice as a function of temperature $T$ and hopping $t = t_1 = t_2$. The phases are labeled as in Figure~\ref{landauPlot}, with `XY MI' referring to a ferromagnetically ordered Mott insulator (MI) where the magnetization vector lies in the $x$-$y$ plane. The two insulating phases are separated by a continuous phase transition (thin line), while the superfluid--insulator transitions are either continuous or first order (thick line). The structure of the phase diagram, including the order of the transitions, agrees with the predictions of Landau theory shown in Figure~\ref{landauPlot}. (At least in this approximation, the transition between the spin-disordered MI and superfluid remains first order for arbitrarily large $t$, indicating that the point $A = \frac{1}{4}$ in the Landau theory is never reached.)}
\end{figure}

\begin{figure}
\putinscaledfigure{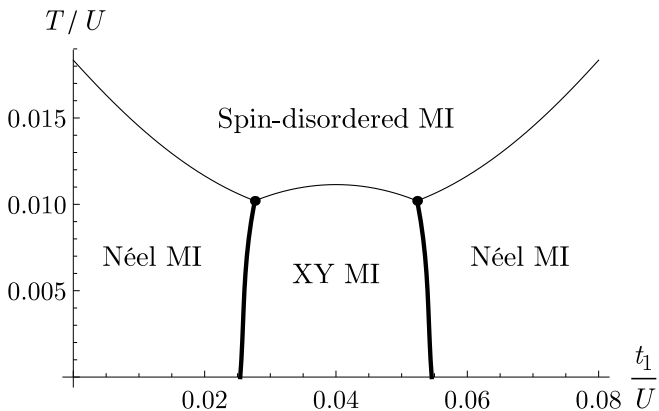}
\caption{\label{Phases3d8}Phase diagram as a function of temperature $T$ and hopping $t_1$, with fixed $t_1 + t_2 = 0.08U$. For this value of the total hopping, there is no superfluid phase. For high temperatures, the system is a Mott insulator (MI) without spin order. For lower temperatures and $t_1 \simeq t_2$, a ferromagnetically ordered state (labeled XY MI) is favored, with the magnetization vector in the $x$-$y$ plane. For more imbalanced hopping values, a N\'eel state is preferred, with a two-sublattice order and the staggered magnetization vector parallel to the $z$ axis. As expected from the Landau theory of Section~\ref{sec:Landau}, the transition from the spin-disordered insulator to either magnetic phase is continuous, but the transition between the two magnetic orderings is of first order.}
\end{figure}

\begin{figure}
\putinscaledfigure{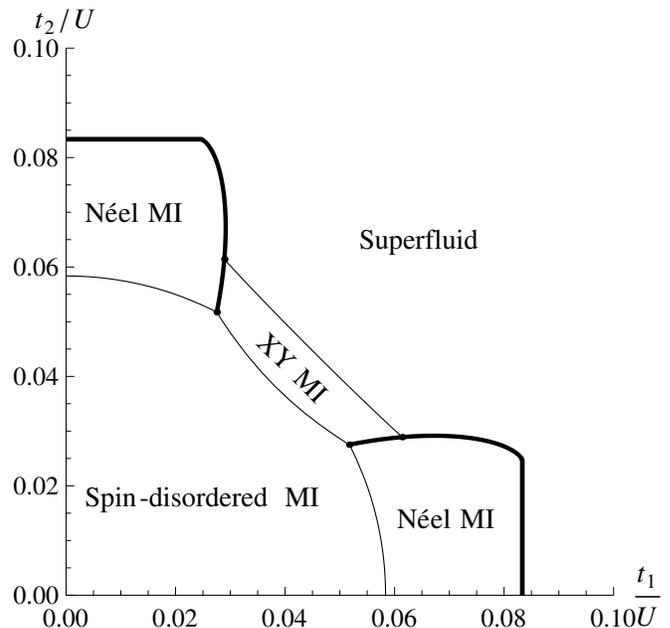}
\caption{\label{Phases3dfixedT}Phase diagram at fixed temperature $T = 0.01 U$, as a function of the two hoppings $t_1$ and $t_2$. At this temperature, all four phases appear for certain values of the hoppings, and the transition between the XY-ferromagnet and the superfluid is continuous. As in Figure~\ref{Phases3d8}, an antiferromagnetically ordered N\'eel state is favored for more imbalanced hoppings. The transition from this phase to the superfluid is of first order, in agreement with the conclusions of Landau theory. This phase diagram is directly comparable to Figure~5 of Ref.~\onlinecite{Altman}, which is plotted at zero temperature, where there is no spin-disordered MI.}
\end{figure}

This approximate treatment allows the phase structure to be found, as a function of the two hopping strengths, $t_1$ and $t_2$, and the temperature $T$, which we measure in units of the interspecies repulsion $U$. Sections through the three-dimensional parameter space are shown in Figures~\ref{Phases3dtT}, \ref{Phases3d8}, and \ref{Phases3dfixedT}, where, as in Figure~\ref{landauPlot}, first-order transitions are indicated by thick lines, and continuous transitions by thin lines. All three phase diagrams have been calculated assuming nearest-neighbor hopping on a simple cubic lattice.

Figure~\ref{Phases3dtT} shows the phase diagram in the plane of equal hopping for the two species, $t_1 = t_2$. As in the perturbative calculation of Section~\ref{sec:PerturbationTheory}, ferromagnetic order is preferred over N\'eel, and the phase diagram contains the XY-ferromagnet, as well as the spin-disordered Mott insulator and the superfluid. The general phase structure is therefore directly comparable to, and in agreement with, that predicted by Landau theory shown in Figure~\ref{landauPlot}.

It should be noted that the maximum temperature for the XY-ferromagnet is limited by the presence of superfluid order. While the transition temperature for magnetic order increases with hopping, as in the perturbative calculation of Section~\ref{sec:PerturbationTheory}, that for superfluidity increases more rapidly. The superfluid replaces the ferromagnetic insulator entirely for moderate $t/U$, at which point the global maximum of the ferromagnetic transition temperature is reached.

Figure~\ref{Phases3d8} shows a phase diagram with fixed $t_1 + t_2 = 0.08U$, as a function of hopping $t_1$ and temperature $T$. Near $t_1 = 0.04U$, where the hopping strengths are equal, the XY-ferromagnet has lower free energy, but for larger disparity in the hoppings, the N\'eel phase is instead favored. As predicted in Section~\ref{sec:Landau}, there is a first-order transition between the two, at which both magnetic order parameters change discontinuously. The value of $t_1$ at this transition line is only weakly dependent on temperature, with the extent of the N\'eel phase increasing slightly with $T$. The two magnetically ordered phases survive only until a temperature of order $1\%$ of $U$, where there is a continuous transition into the magnetically disordered phase. (An attempt to raise the critical temperature further by increasing the hopping leads instead to the superfluid phase.)

In Figure~\ref{Phases3dfixedT}, the phase diagram is shown for fixed temperature $T = 0.01 U$, with all four phases appearing for suitable choices of the hopping strengths $t_1$ and $t_2$. At this temperature, the transition from the ferromagnetic insulator to the superfluid is continuous, but, as predicted in Section~\ref{sec:Landau}, the transition from the N\'eel phase to the superfluid is of first order.

\section{Discussion}
\label{sec:Conclusions}

This work has presented a study of a system of two species of bosons moving in a lattice, based on two approaches. First, the standard techniques of Landau theory were used to determine the nature of the possible phase boundaries. Then an analytic framework was introduced, based on a field-theoretical representation of the partition function. This builds upon the standard mean-field theory for the spinless Bose-Hubbard model,\cite{FWGF,QPT} but takes into account the important effects of fluctuations, leading to magnetic order. Extensions to the method could be used to describe other bosonic models and to calculate other properties, such as dynamic correlations.

An important conclusion of this work is that the maximum temperature at which magnetically ordered insulating phases can survive is determined by the presence of the superfluid phase. While the critical temperature for magnetic ordering increases with hopping, in line with the predictions of a simple perturbative calculation, the same is true of the superfluid transition temperature. For moderate values of the hopping, reducing the temperature from the spin-disordered Mott insulator leads directly to the superfluid phase, without passing through a magnetically ordered insulator (see Figure~\ref{Phases3dtT}).

This implies that an experimental realization of the magnetically ordered insulators will require temperatures in the lattice on the order of $1\%$ of the on-site repulsion $U$. At these temperatures, thermal particle-number fluctuations are completely frozen out, and the only entropy results from spin fluctuations. Reaching these phases will therefore likely require a considerable reduction in the entropy after loading into the lattice, using techniques such as those proposed by Bernier et al.\cite{Bernier} To give precise predictions for the temperature and entropy constraints, further numerical work is needed, and the effects of the trapping potential, which has been neglected here, should be taken into account.

As pointed out by Altman et al.,\cite{Altman,Altman2} the magnetically ordered phases could be detected by analyzing correlations of the shot noise in time-of-flight measurements. It would also be interesting to use recently developed techniques of photoemission spectroscopy for cold atoms\cite{Stewart} to investigate the effect of the magnetic ordering transition on the single-particle Green function.\cite{ExcitedStateSpectra}

\begin{acknowledgments}
I thank J. T. Chalker and F. H. L. Essler for helpful discussions. The work was supported by EPSRC Grant No.\ EP/D050952/1.
\end{acknowledgments}

\end{document}